\documentclass[prd,aps,12pt,nofootinbib,superscriptaddress]{revtex4-1}

\usepackage{amsmath}
\usepackage{graphicx}

\newcommand{\beq}{\begin{equation}}
\newcommand{\eeq}{\end{equation}}

\newcommand{\ber}{\begin{eqnarray}}
\newcommand{\eer}{\end{eqnarray}}

\def\rb{{\rm b}}

\def\m{{\rm m}}

\def\C{{\cal C}}

\def\beb{{b}}
\def\muk{{\Theta}}
\def\l{\Lambda}

\begin{document}

\title{A no-boundary proposal for braneworld perturbations}

\author{Alexander Viznyuk}
\affiliation{Bogolyubov Institute for Theoretical Physics, Kiev 03680, Ukraine} %
\affiliation{Department of Physics, Taras Shevchenko Kiev National University, Kiev, Ukraine} %

\author{Yuri Shtanov}
\affiliation{Bogolyubov Institute for Theoretical Physics, Kiev 03680, Ukraine} %
\affiliation{Department of Physics, Taras Shevchenko Kiev National University, Kiev, Ukraine} %

\author{Varun Sahni}
\affiliation{Inter-University Centre for Astronomy and Astrophysics, Pune, India} %

\begin{abstract}
We propose a novel approach to the problem of cosmological perturbations in a braneworld
model with induced gravity, which leads to a closed system of equations on the brane.  We
focus on a spatially closed brane that bounds the interior four-ball of the bulk space.
The background cosmological evolution on the brane is now described by the normal branch,
and the boundary conditions in the bulk become the regularity conditions for the metric
everywhere inside the four-ball. In this approach, there is no spatial infinity or any
other boundary in the bulk space since the spatial section is compact, hence, we term
this setup as a {\em no-boundary\/} proposal. Assuming that the bulk cosmological
constant is absent and employing the Mukohyama master variable, we argue that the effects
of nonlocality on brane perturbations may be ignored if the brane is marginally closed.
In this case, there arises a relation that {\em closes the system of equations\/} for
perturbations on the brane. Perturbations of pressureless matter and dark radiation can
now be described by a system of coupled second-order differential equations. Remarkably,
this system can be {\em exactly solved\/} in the matter-dominated and de~Sitter regimes.
In this case, apart from the usual growing and decaying modes, we find two additional
modes that behave monotonically on super-Hubble spatial scales and exhibit rapid
oscillations with decaying amplitude on sub-Hubble spatial scales.
\end{abstract}

\maketitle

\tableofcontents

\section{Introduction}\label{sec: intro}

In the braneworld paradigm \cite{Maartens:2010ar}, our observable universe is a
four-dimensional hypersurface (the ``brane'') embedded in a higher-dimensional spacetime
(the ``bulk'') with Standard-Model particles and fields trapped on the brane. In the most
popular cosmological models, there is only one extra dimension, and the brane is either a
boundary or is embedded into a five-dimensional bulk space-time.  From the
four-dimensional viewpoint, this can be regarded as a modification of gravity.  In one of
its popular versions, the Randall--Sundrum (RS) model \cite{RS}, general relativity (and
the inverse square law) is modified due to extra-dimensional effects on relatively small
spatial scales. Apart from interesting high-energy implications of this model, it was
shown to be potentially capable of explaining the observations of the galactic rotation
curves and X-ray profiles of galactic clusters without invoking the notion of dark matter
\cite{Rotation curves}.

In another version of the braneworld paradigm, first proposed by Arkani-Hamed, Dimopoulos
and Dvali \cite{ArkaniHamed:1998rs} and subsequently developed by Dvali, Gabadadze and
Porrati \cite{DGP} (the DGP model), gravity is modified on large spatial scales. This
model is based on the inclusion of the Hilbert--Einstein term in the action for the brane
(so-called induced gravity), and, because of this, has two branches of cosmological
solutions. The `self-accelerating' branch can model cosmology with late-time acceleration
without cosmological constants either in the bulk or on the brane, while the `normal'
branch requires at least a brane tension to accelerate the expansion. Subsequent analysis
has revealed a serious stability issue of the self-accelerating branch of the DGP model
--- the so-called ghost instability \cite{Ghosts}.  This leaves the normal branch,
perhaps, as the only physically viable solution, consistent with the current observation
of cosmological acceleration.  A general braneworld model containing the induced gravity
term as well as cosmological constants in the bulk and on the brane, was first proposed
and studied in \cite{Collins:2000yb, Shtanov:2000vr, Deffayet:2000uy}. In describing the
late-time cosmological acceleration on the normal branch, this model exhibits a number of
interesting generic features, including the possibility of superacceleration
(supernegative, or phantom-like, effective equation of state of dark energy $w_{\rm eff}
\leq -1$) \cite{Sahni:2002dx, Lue:2004za}. It is interesting that a phantom-like equation
of state arises in the normal branch {\em without\/} the development of a `Big-rip'
future singularity. Such an effective equation of state appears to be consistent with the
most recent set of observations of type Ia supernovae combined with other data sets
\cite{Rest:2013bya}. Such a braneworld also admits the possibility of cosmological
loitering even in a spatially flat universe \cite{Sahni:2004fb}, and the property of
cosmic mimicry, wherein a low-density braneworld shares the {\em precise\/}
 expansion history of $\Lambda$CDM \cite{Sahni:2005mc};
for a brief review see \cite{Sahni:review}. This braneworld model can also be used to
address astrophysical observations of dark matter in galaxies \cite{Viznyuk:2007ft}.

The theory of structure formation, temperature anisotropy of the cosmic microwave
background (CMB) and other issues that form the basis of experimental tests of any
cosmological model require the knowledge of the evolution of cosmological perturbations.
Developing the theory of cosmological perturbations in the braneworld context is a
long-standing problem. The existence of an extra dimension requires taking into account
the corresponding dynamical degrees of freedom and the specification of the boundary
conditions in the bulk space.  In the case of a spatially flat brane, the extra dimension
is noncompact, and one has to deal with its spatial infinity. The main difficulty, in
this case, is the presence of the bulk gravitational effects which lead to the
non-locality of the resulting equations on the brane. In spite of the analytical
complexity of the problem, considerable progress has been made in this direction over the
last few years.  Based on a very convenient Mukohyama master variable and master equation
\cite{Mukohyama:2000ui, Mukohyama:2001yp}, interesting results were presented in
\cite{Deffayet:2002fn, Deffayet:2004xg, Koyama:2005kd, Koyama:2006ef, Sawicki:2006jj,
Song:2007wd, Seahra:2010fj}. However, all {\em analytical\/} results (of which we are
aware) are based on some kind of simplifying assumptions or approximations which are
taken for granted. Thus, the well-known quasi-static (QS) approximation introduced by
Koyama and Maartens \cite{Koyama:2005kd} (for an extension into the non-linear regime,
see \cite{non-linear}) is based on the assumption of {\em slow temporal evolution\/} of
all quantities on sub-Hubble spatial scales, as compared to spatial gradients. Another
approximation --- the dynamical scaling (DS) ansatz, proposed by Sawicki {\em et al.\@}
\cite{Sawicki:2006jj, Song:2007wd} --- assumes that perturbations evolve as a power of
the scale factor with time-varying index. Both approximations were further analyzed in
\cite{Seahra:2010fj}. A complete system of equations in the bulk and on the brane were
solved numerically in the framework of the RS \cite{Cardoso:2007zh} and DGP
\cite{Cardoso:2007xc} braneworld models.

Notwithstanding the partial success of approximate methods and numerical integration
schemes, the problem of cosmological perturbations on the brane cannot be viewed as
having been successfully solved without either a justification of the introduced
approximations or an analytical solution. In this work, we propose a new approach to the
issue of boundary conditions in the bulk which permits one to obtain a closed system of
equations for scalar cosmological perturbations on a (marginally) closed brane without
any simplifying assumptions (aside from assuming a small spatial curvature for the
brane).

\begin{figure}[htb]
\begin{center}
\includegraphics[width=.25\textwidth]{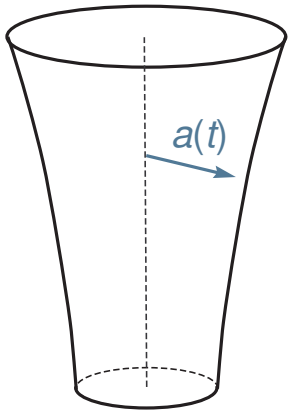}
\caption{The spatially closed brane has a three-sphere ($S^3$) topology and bounds a
four-ball in the bulk space. The expansion factor, $a(t)$, of the brane corresponds to
the radial coordinate of the bulk space, as shown in this figure and discussed in greater
detail in Sec.~\ref{sec: equations in the flat background bulk geometry}.
\label{fig:brane}}
\end{center}
\end{figure}

In the case of a spatially flat brane, the `standard' method of setting boundary
conditions on the past Cauchy horizon does not guarantee the absence of a singularity in
the whole of the bulk space (for a discussion of this problem, see
\cite{Shtanov:2007dh}). In the present paper, we consider a spatially {\em closed\/}
brane, which is a boundary of the four-ball of the bulk space. (Alternatively, one can
picture the spatially closed brane as embedded in the bulk with $Z_2$ symmetry of
reflection with respect to the brane; in this case, there are two identical bulk spaces
with four-ball topology on the two sides of the brane.) This configuration gives the
normal branch of the theory and describes the model originally suggested in
\cite{Shtanov:2000vr}. There is no spatial infinity (or another boundary), and the
boundary conditions in the bulk become natural and simple: the metric is required to be
regular everywhere inside of the (evolving) four-ball bounded by the brane (see
Fig.~\ref{fig:brane}). In this regard, we call this setup a {\em no-boundary\/} proposal.
In the simplest case where the unperturbed bulk space is flat (the bulk cosmological
constant is equal to zero), we succeed in deriving a {\em closed\/} approximate system
for scalar perturbations on the brane and find some exact analytical solutions.

The most important qualitative result of this investigation is that, apart from the usual
monotonic growing and decaying modes for scalar-type perturbations, we find another two
set of modes which behave in an oscillatory fashion with a slowly decaying amplitude.

Our paper is organized as follows. In the next section, we introduce the braneworld equations and
discuss their (unperturbed) cosmological solutions. In Sec.~\ref{sec: perturbations on the
brane}, we present the system of equations describing scalar cosmological perturbations
on the brane. This system is not closed because the evolution equation for
the anisotropic stress of the bulk projection (the so-called Weyl fluid, or dark
radiation) is missing. The anisotropic stress of the Weyl fluid is then expressed through
the Mukohyama master variable in Sec.~\ref{sec: equations in the flat background bulk
geometry} for the case of a spatially closed brane which bounds the bulk space.
In this case, a general solution for the master variable
can be found. In Sec.~\ref{sec: perturbations on a marginally closed brane}, we prove
that the effects of nonlocality may be ignored if the brane is marginally closed, and
obtain a closed system of second-order differential equations describing the
perturbations of pressureless matter and the Weyl fluid in this case. Some exact solutions of
these equations are presented in Sec.~\ref{sec: general-relativistic regime}. The results
of our work are summarized in Sec.~\ref{sec: conclusion}.

\section{Background cosmological evolution}\label{sec: background}

We start from the generic form of the braneworld action:
\beq \label{action}
S = M^3 \left[\int_{\rm bulk} \left( {\cal R} - 2 \Lambda \right) - 2 \int_{\rm brane} K
\right] +  \int_{\rm brane} \left( m^2 R - 2 \sigma \right) + \int_{\rm brane} L \left(
g_{\mu\nu}, \phi \right) \, ,
\eeq
where ${\cal R}$ is the scalar curvature of the five-dimensional metric $g_{AB}$, and $R$
is the scalar curvature corresponding to the induced metric $g_{\mu\nu}$ on the
brane.\footnote{Here and below, we use upper-case Latin indices $A,B,\ldots$ for the
five-dimensional bulk coordinates and Greek indices $\mu,\nu,\ldots$ for the
four-dimensional coordinates on the brane.} The symbol $L \left( g_{\mu\nu}, \phi
\right)$ denotes the Lagrangian density of the four-dimensional matter fields $\phi$
whose dynamics is restricted to the brane so that they interact only with the induced
metric $g_{\mu\nu}$. The quantity $K$ is the trace of the symmetric tensor of extrinsic
curvature of the brane. All integrations over the bulk and brane are taken with the
corresponding natural volume elements. The universal constants $M$ and $m$ play the role
of the five-dimensional and four-dimensional Planck masses, respectively. The symbol
$\Lambda$ denotes the bulk cosmological constant, and $\sigma$ is the brane tension.

The actions of the RS \cite{RS} and DGP \cite{DGP} braneworld models are special cases of
(\ref{action}). The action of the RS braneworld is obtained after setting $m=0$ in
\eqref{action}, whereas neglecting $\Lambda$ and $\sigma$ leads
to
the DGP model. General relativity with $1/m^2$ playing the role of the
gravitational constant can also be formally obtained from \eqref{action} after setting $M
= 0$.

The action \eqref{action} leads to the Einstein equation with a
cosmological constant in the bulk,
\begin{equation} \label{bulk}
{\cal G}_{AB} + \Lambda g_{AB} = 0 \, ,
\end{equation}
and the following effective equation on the brane \cite{Shiromizu:1999wj, Sahni:2005mc}:
\begin{equation}
  G_{\mu\nu} + \left(\frac{\Lambda_{\rm RS}}{\beb + 1}\right) g_{\mu\nu}
 = \left(\frac{\beb}{\beb + 1}\right) \frac{1}{m^2} \, T_{\mu\nu} +
\left(\frac{1}{\beb + 1}\right) \left[ \frac{1}{M^6} Q_{\mu\nu} -
\C_{\mu\nu} \right] \, ,\label{effective} \end{equation} where
\begin{equation}\label{beta}
\beb = k \ell\,, \qquad k=\frac{\sigma}{3M^3}\,, \qquad
\ell=\frac{2m^2}{M^3}\,,\qquad \Lambda_{\rm
RS}=\frac{\Lambda}{2}+\frac{\sigma^2}{3M^6}
\end{equation}
are convenient parameters to describe the braneworld model, and
\begin{equation}\label{q}
Q_{\mu\nu} = \frac13 E E_{\mu\nu} - E_{\mu\lambda} E^{\lambda}{}_\nu + \frac12
\left(E_{\rho\lambda} E^{\rho\lambda} - \frac13 E^2 \right) g_{\mu\nu}\,,
\end{equation}
\begin{equation}\label{bare einstein}
E_{\mu\nu} \equiv m^2 G_{\mu\nu} - T_{\mu\nu} \,.
\end{equation}

Gravitational dynamics on the brane is not closed because of the presence of the
symmetric traceless tensor $\C_{\mu\nu}$ in \eqref{effective}, which is the projection of
the five-dimensional Weyl tensor from the bulk onto the brane. The tensor $\C_{\mu\nu}$
is not freely specifiable on the brane, but is related to the tensor $Q_{\mu\nu}$ through
the conservation equation
\begin{equation}\label{conserv_weyl}
\nabla^\mu \left( Q_{\mu\nu} - M^6 \C_{\mu\nu} \right) = 0 \, ,
\end{equation}
which is a consequence of the Bianchi identity applied to \eqref{effective} and the law
of stress--energy conservation for matter:
\begin{equation}\label{conserv}
\nabla^\mu T_{\mu\nu} = 0 \, .
\end{equation}
Here, $\nabla_\mu$ denotes the derivative on the brane compatible with the induced metric
on the brane.

The cosmological evolution of the Friedmann--Robertson--Walker (FRW) braneworld model
\begin{equation}
\label{FRW} ds^2 = - d t^2 + a^2 (t) \gamma_{ij} d x^i dx^j \,
\end{equation}
can be obtained from \eqref{effective} with the following result \cite{Collins:2000yb,
Shtanov:2000vr, Deffayet:2000uy, Sahni:2002dx}:
\begin{equation} \label{background}
H^2 + {\kappa \over a^2} = {\rho + \sigma \over 3 m^2} + {2 \over \ell^2}\left[ 1 \pm
\sqrt{1 + \ell^2 \left({\rho + \sigma \over 3 m^2} - {\Lambda \over 6} - {C \over a^4}
\right)} \right] \, .
\end{equation}
Here, $H\equiv \dot{a}/a$ is the Hubble parameter, $\rho=\rho(t)$ is the energy density of
matter on the brane and $C$ is a constant resulting from the presence of the symmetric
traceless tensor $\C_{\mu\nu}$ in the field equations \eqref{effective}. The parameter
$\kappa = 0, \pm 1$ corresponds to different spatial geometries of the maximally
symmetric spatial metric $\gamma_{ij}$.

The $\pm$ sign in equation \eqref{background} reflects the two different ways in which
the bulk can be bounded by the brane \cite{Deffayet:2000uy}, resulting in two different
branches of solutions. These are usually called the normal branch ($-$ sign) and
the self-accelerating branch (+ sign).

\section{Scalar cosmological perturbations: combined system of equations on the brane}
\label{sec: perturbations on the brane}

The scalar cosmological perturbations of the induced metric on the brane are most
conveniently described by the relativistic potentials $\Phi$ and $\Psi$ in the
longitudinal gauge:
\begin{equation} \label{metric}
ds^2 = - ( 1 + 2 \Phi) d t^2 + a^2 ( 1 - 2 \Psi) \gamma_{ij} d x^i dx^j \, .
\end{equation}

The components of the linearly perturbed stress--energy tensor of matter in these
coordinates are defined as follows\footnote{The spatial indices $i,j,\ldots$ in purely
spatially defined quantities (such as $v_i$ and $\delta \pi_{ij}$) are always raised and
lowered using the spatial metric $\gamma_{ij}\,$; in particular, $\gamma^i{}_j =
\delta^i{}_j$. The symbol $\nabla_i$ denotes the covariant derivative with respect to the
spatial metric $\gamma_{ij}$, and the spatial Laplacian is $\nabla^2 = \nabla^i
\nabla_i$.}:
\begin{equation}
\delta T^\mu{}_\nu = \left(
\begin{array}{cc}
\displaystyle - \delta \rho \, , & \displaystyle - \nabla_i v
\medskip \\
\displaystyle  {\nabla^i v \over a^2} \, , & \displaystyle \delta
p\, \delta^i{}_j + \frac{\zeta^i{}_j}{a^2}
\end{array}
\right) \, ,
\end{equation}
where $\delta \rho$, $\delta p$, $v$, and $\zeta_{ij} = \left( \nabla_i \nabla_j -
\frac13 \gamma_{ij} \nabla^2 \right) \zeta$ describe the scalar perturbations.

Similarly, we introduce the scalar perturbations $\delta \rho_\C$,
$v_\C$, and $\delta \pi_\C$ of the tensor $\C_{\mu\nu}$:
\begin{equation} \label{weyl projection definition}
m^2 \delta \C^\mu{}_\nu = \left(
\begin{array}{cc}
\displaystyle - \delta \rho_\C
\, , & \displaystyle - \nabla_i v_\C  \medskip \\
\displaystyle {\nabla^i v_\C \over a^2} \, , & \displaystyle {\delta
\rho_\C \over 3 } \delta^i{}_j + {\delta \pi^i{}_j \over a^2}
\end{array}
\right) \, ,
\end{equation}
where $\delta \pi_{ij} = \left( \nabla_i \nabla_j - \frac13
\gamma_{ij} \nabla^2 \right) \delta \pi_\C$.

We call $v$ and $v_\C$ the momentum potentials for matter and dark
radiation, respectively, $\delta \rho$ and $\delta \rho_\C$ are
their energy density perturbations, and $\zeta$ and $\delta\pi_\C$
are the scalar potentials for their anisotropic stresses.

Using this notation, one can derive a complete system of equations that describe the
evolution of scalar cosmological perturbations on the brane (for details of the
derivation see \cite{Shtanov:2007dh, Viznyuk:2012oda}):
\begin{equation}
{1 \over a^2} \left(\nabla^2 + 3 \kappa \right) \Psi = \left(1 + {2 \over \beta} \right)
{\rho\,\Delta \over 2 m^2}  + {\left(\delta \rho_\C + 3 H v_\C \right) \over m^2 \beta }
\, , \label{con-delta1}
\end{equation}
\begin{equation}
m^2 \beta \left( \dot \Psi + H \Phi \right) = \left(1 + {\beta \over 2} \right) v + v_\C
\, , \label{con-v1}
\end{equation}
\begin{eqnarray} \nonumber
\ddot \Psi &+& 3(1 + \gamma) H \dot \Psi + H \dot \Phi + \left[ 2 \dot H + 3 H^2 (1 +
\gamma) \right] \Phi \,- \\ \nonumber &-& {\gamma \over a^2} \nabla^2 \Psi - {\kappa (1 +
3 \gamma) \over a^2} \Psi
 + {1 \over 3 a^2} \nabla^2 (\Phi - \Psi) \,= \\ &=& \displaystyle - \left(
\gamma + {2 \over 3 \beta} \right) {\delta \rho \over 2 m^2 } + \left( 1 + {2 \over
\beta} \right) {\delta p \over 2 m^2} \, , \label{psi1}
\end{eqnarray}
\begin{equation}
\frac{\delta \pi_\C}{m^2} = - \frac{\beta (3 \gamma + 1)}{4} \left( \Phi - \Psi + {\zeta
\over m^2} \right) - \frac{\zeta}{m^2} \, , \label{pi1}
\end{equation}
\begin{equation}
\delta \dot \rho + 3 H ( \delta \rho + \delta p ) = {1 \over a^2} \nabla^2 v + 3 (\rho +
p ) \dot \Psi \, , \label{m11}
\end{equation}
\begin{equation}
\dot v + 3 H v = \delta p + (\rho + p) \Phi + { 2 \over 3 a^2} \left( \nabla^2 + 3 \kappa
\right) \zeta \, , \label{m22}
\end{equation}
\begin{equation}
\delta \dot \rho_\C + 4 H \delta \rho_\C = {1 \over a^2} \nabla^2 v_\C - {12 m^2 C \over
a^4} \dot \Psi \, , \label{w11}
\end{equation}
\begin{eqnarray}
\dot v_\C &+& 3 H v_\C = \frac13 \delta \rho_\C - {4 m^2 C \over a^4 } \Phi + \frac16
\beta (1 - 3 \gamma) \rho\, \Delta -  \nonumber \\ &-& {2 + \beta \over 3 a^2}
\left(\nabla^2 + 3 \kappa \right) \zeta - {m^2 \beta \over 3 a^2} \left(\nabla^2 + 3
\kappa \right) \left[ \Phi - 3 \gamma \Psi \right]\, ,  \label{w33}
\end{eqnarray}
where
\begin{equation}\label{Delta}
\Delta \equiv \frac{\delta \rho + 3 H \upsilon}{\rho}
\end{equation}
is the gauge-invariant density perturbation, and the time-dependent dimensionless
functions $\beta$ and $\gamma$ are given by
\begin{equation} \label{beta-gamma}
\beta \equiv \pm \, 2 \sqrt{1 + \ell^2 \left({\rho + \sigma \over 3 m^2} -
{\Lambda \over 6} - {C \over a^4} \right) } \, , \qquad \gamma \equiv \frac13 \left(1 +
{\dot
\beta \over H \beta} \right) \, .
\end{equation}
Equations \eqref{con-delta1}--\eqref{pi1} are the consequence of the field equations
\eqref{effective}, while equations \eqref{m11}--\eqref{w33} are the result of the
conservation laws \eqref{conserv_weyl}, \eqref{conserv}. Note that the upper sign in
$\beta$ corresponds to the self-accelerating branch (of the braneworld) while the lower
sign leads to the normal branch. It is well known that the self-accelerating branch is
plagued by ghosts \cite{Ghosts}.  We do not encounter this problem here because only the
normal branch is discussed in this paper, due to the no-boundary setup described in the
introduction.

If dark radiation is absent ($C = 0$), then from (\ref{con-delta1})--(\ref{w33}) one can
derive the following useful system for perturbations in {\em pressureless\/} matter ($p =
0$, $\zeta = 0$):
\begin{equation}
\ddot \Delta + 2 H \dot \Delta = \left(1 + {6 \gamma \over
\beta} \right) {\rho \Delta \over 2 m^2} + (1 + 3 \gamma)
{\delta\rho_\C \over m^2 \beta}  \, , \label{one}
\end{equation}
\begin{equation}
\delta \dot \rho_\C + 4 H \delta \rho_\C = {1 \over a^2} \nabla^2 v_\C \label{two} \, ,
\end{equation}
\begin{equation}
\dot v_\C + H (4-3\gamma) v_\C = \gamma \delta \rho_\C + \left(\gamma - \frac13 \right)
\rho\,\Delta \,+ {4 \over 3 (1 + 3 \gamma) a^2 } \left( \nabla^2 + 3 \kappa \right)
\delta \pi_\C \, . \label{three}
\end{equation}

As compared to its counterpart in general relativity,
\begin{equation}
\ddot \Delta + 2 H \dot \Delta = {\rho \Delta \over 2 m^2} \, , \label{gr}
\end{equation}
equation \eqref{one} has three distinctive features that affect the evolution of matter
perturbations:
\begin{itemize}
\item the evolution of the Hubble parameter $H(t)$ is modified by virtue of
    (\ref{background});
\item the effective gravitational constant is renormalized by the time-dependent
    factor $ {(1 + 6 \gamma / \beta)}$;
\item perturbations of the Weyl fluid are present on the right-hand side of
    (\ref{one}) and therefore influence the growth of $\Delta$.
\end{itemize}

Moreover, in contrast to general relativity, the relativistic potentials $\Phi$ and
$\Psi$ are no longer equal to each other, their difference being determined by equation
\eqref{pi1}:
\beq \label{difference}
\Phi - \Psi = - \frac{4 \delta \pi_\C}{\beta (3 \gamma + 1) m^2} \, .
\eeq

Determining the influence of the Weyl fluid on matter perturbations is not an easy task.
In fact, one notes that the system of equations (\ref{one})--(\ref{three}), as well as
(\ref{con-delta1})--(\ref{w33}), is not closed because the evolution equation for the
anisotropic stress (of dark radiation) $\delta \pi_\C$ is missing. Perhaps, the simplest
way to get rid of this problem would be to set $\delta \pi_\C=0$. This type of boundary
condition was shown to be consistent with conservation equations and lead to a closed
system of differential equations describing the evolution of cosmological perturbations
on the brane \cite{Shtanov:2007dh}. However, this way of imposing boundary conditions
directly on the brane, though simple, is not well motivated from the bulk perspective.
Physically, the evolution of the Weyl tensor should be derived from the perturbed bulk
equation \eqref{bulk} after setting some natural boundary conditions in the bulk
\cite{Mukohyama:2000ui, Mukohyama:2001yp, Deffayet:2002fn, Deffayet:2004xg,
Cardoso:2007zh, Cardoso:2007xc, Seahra:2010fj, Koyama:2005kd, Koyama:2006ef,
Sawicki:2006jj, Song:2007wd, Maartens:2010ar}.

In the following, this idea will be realized for a spatially closed ($\kappa=1$)
braneworld model which bounds the interior bulk space with the spatial topology of a ball
(see Fig.~\ref{fig:brane}). The background cosmological evolution on the brane in this
case is described by the ghost-free normal branch, and the boundary conditions in the
bulk take the form of a regularity condition inside the four-ball. We will show in the
next section that, if the background bulk metric is flat,\footnote{This deviates from the
original DGP model \cite{DGP} only by a non-vanishing brane tension and closed spatial
geometry of the brane. On the other hand, this is the model originally considered in
\cite{Shtanov:2000vr} but with the zero cosmological constant in the bulk.} then the
perturbed bulk equations for such a configuration can be solved explicitly.

\section{Perturbations around a flat bulk geometry}
\label{sec: equations in the flat background bulk geometry}

We noted in the previous section that, in order to close the system of equations
(\ref{one})--(\ref{three}), one required an additional evolution equation for the
anisotropic stress of dark radiation $\delta \pi_\C$. Such an equation can be derived by
considering perturbations in the bulk.

To begin with, note that a flat background bulk metric (which requires $C=0$ and
$\Lambda=0$) bounded by a spatially closed ($\kappa=1$) brane is most conveniently
described in the natural static coordinates
\begin{equation} \label{bmetric}
ds^2_{\rm bulk} = - d \tau^2 +d r^2 + r^2 \gamma_{ij} dx^i dx^j \, ,
\end{equation}
where $\gamma_{ij}$ is the metric of a maximally symmetric space with coordinates $x^i$,
the same as in \eqref{FRW}. In these coordinates, the FRW brane moves radially along the
trajectory $r = a(t)$, $\tau=\tau(t)$, and the relevant part of the bulk is given by $0
\leq r \leq a(t)$. The parameter $t$, which describes the cosmological time on the brane,
is the same as in \eqref{FRW}. The quantity $a(t)$ is the scale factor of the background
Friedman--Robertson--Walker metric on the brane, which evolves according to
\eqref{background}, and the function $\tau(t)$ is defined by the differential equation $d
\tau / d t = \sqrt{1+a^2 H^2}$, where $H=\dot{a}/a$ is the Hubble parameter on the brane.
An overdot in this paper denotes differentiation, with respect to $t$, of quantities defined
on the brane.

In the following, we shall adopt the standard decomposition in terms of three-spherical
harmonics with respect to the coordinates $x^i$ in (\ref{bmetric}).  For any scalar
quantity $f (\tau, r, x^i)$ defined in the bulk, one has
\beq
f (\tau, r, x^i) = \sum_{n l m} f_{n l m} (\tau, r) Y_{n l m} (x^i) \, ,
\eeq
where $Y_{n l m} (x^i)$, $m = -l, -l + 1, \ldots, l$, $l = 0, \ldots, n$, $n = 0, 1, 2,
\ldots$, is the orthonormal system of scalar harmonics on the unit three-sphere which are
eigenfunctions of the Laplace operator:
\beq
\nabla^2 Y_{n l m} (x^i) = - n (n + 2) Y_{n l m} (x^i) \, .
\eeq
Similarly, any scalar function on the brane is expanded in terms of scalar harmonics
$Y_{n l m} (x^i)$ with respect to the coordinates $x^i$.  In proceeding to the harmonic
coefficients, the three-spherical harmonic indices $n$, $l$, $m$ will be suppressed in
this paper.

Scalar perturbations of the Weyl fluid can be expressed in terms of the Mukohyama master
variable \cite{Mukohyama:2000ui, Mukohyama:2001yp} $\muk(\tau,r)$, defined in the bulk,
as follows (see \cite{Viznyuk:2012oda} for details):
\begin{equation}
\frac{\delta \rho_{\C}}{m^2} = -\,\frac{n (n+2)(n^2+2n-3)}{3 a^5}\,\muk_\rb \,,
\label{weyl projection rho master}
\end{equation}
\begin{equation}
\frac{v_{\C}}{m^2} = \,\frac{(n^2+2n-3)}{3 a^3} \left[ a H \left(\partial_{r}
\muk\right)_\rb + \sqrt{1 + a^2 H^2}\left(\partial_{\tau} \muk\right)_\rb -
H\muk_\rb\right] \,, \label{weyl projection upsilon master}
\end{equation}
\begin{eqnarray}
\frac{\delta \pi_{\C}}{m^2} &=& - \,\frac{ \left(1 + 2 a^2
H^2\right)}{2a}\left(\partial^2_{\tau}\muk\right)_\rb\! -
\frac{\left(1+3 a^2 H^2\right)}{2a^2}\left(\partial_r \muk\right)_\rb \!- \nonumber \\
&-&  H \sqrt{1 + a^2 H^2}\left(\partial^2_{\tau r} \muk\right)_\rb  \, -
\frac{(n^2+2n-3)\left(1+3 a^2 H^2\right)}{6a^3}\, \muk_\rb \,, \label{weyl projection pi
master}
\end{eqnarray}
where the subscript $\{\}_\rb$ means that the corresponding quantity is evaluated at the
brane; for example, $\muk_\rb (t) \equiv \muk \left( \tau(t), a(t) \right)$.  For the
case of a spatially flat brane, equations \eqref{weyl projection rho master}--\eqref{weyl
projection pi master} were previously derived in \cite{Deffayet:2002fn}.

Using the rule of differentiation
\begin{equation} \label{rule of differentiation}
\dot{\muk}_\rb=\sqrt{1+a^2 H^2}\left(\partial_{\tau} \muk\right)_\rb + a H
\left(\partial_{r} \muk\right)_\rb \, ,
\end{equation}
one can write \eqref{weyl projection upsilon master}, \eqref{weyl projection pi master}
in the following form:
\begin{equation}
\frac{v_{\C}}{m^2} = \,\frac{(n^2+2n-3)}{3 a^3} \left[ \dot{\muk}_\rb - H\muk_\rb\right]
\,, \label{weyl projection upsilon master t}
\end{equation}
\begin{equation}
\frac{\delta \pi_{\C}}{m^2} = - \,\frac{1}{2a} \left[\ddot{\muk}_\rb - \frac{a^2 H
(\dot{H}+H^2)}{(1 + a^2 H^2)}\,\dot{\muk}_\rb +  \frac{(1 - a^2 \dot{H})}{a(1 + a^2
H^2)}\left(\partial_r \muk\right)_\rb + \frac{(n^2+2n-3)}{3a^2}\, \muk_\rb \right] \, .
\label{weyl projection pi master t}
\end{equation}
Using \eqref{weyl projection rho master} and \eqref{weyl projection upsilon master t},
one can relate the functions $v_\C(t)$ and $\delta\rho_\C (t)$ as follows:
\begin{equation} \label{upsilon C}
v_\C = -\, \frac{a^2}{n(n+2)} \left(\delta \dot \rho_\C + 4 H \delta \rho_\C \right)\, .
\end{equation}
This is in accordance with one of the conservation equations \eqref{w11} on the brane,
(recall that $C = 0$ in our case).

However, relating $\delta\pi_\C (t)$ and $\delta\rho_\C (t)$ is not so trivial because of
the term proportional to $\left(\partial_r \muk\right)_\rb$ in \eqref{weyl projection pi
master t}. To establish a relation between $\delta\pi_\C$ and $\delta\rho_\C $
one needs to determine the Mukohyama master variable in the bulk. This is what we proceed
to do next.

Disregarding the trivial modes with $n = 0, 1$, we have the following equation for the
Mukohyama master variable (see \cite{Mukohyama:2000ui}):
\begin{equation} \label{master equation flat}
-\partial^{2}_{\tau} \muk+\partial^{2}_{r} \muk-\frac{3}{r}
\partial_{r} \muk-\frac{(n^2+2n-3)}{r^2}\, \muk=0 \, .
\end{equation}
This is a partial differential equation of hyperbolic type. In the coordinates we have
chosen, it has quite a simple form, allowing one to separate the variables by setting
$\muk(\tau,\,r)=\xi(\tau)\,\chi(r)$ with the functions $\xi(\tau)$ and $\chi(r)$
satisfying the ordinary differential equations
\begin{equation} \label{separation variables 1} \frac{d^2\xi(\tau)}{d\tau^2}+W\xi(\tau)=0\, ,
\end{equation}
\begin{equation} \label{separation variables 2}
\frac{d^2\chi(r)}{dr^2} - \frac{3}{r} \frac{d\chi(r)}{dr} + \left[W -
\frac{(n^2+2n-3)}{r^2} \right] \chi(r)=0 \, ,
\end{equation}
where $W$ is some constant. Depending on the sign of the constant $W$, we have two
qualitatively different solutions of \eqref{separation variables 1}, either with
oscillatory or with exponential behavior. We consider them separately.

Setting $W\equiv\omega^2 > 0$, we get the solution of \eqref{separation variables 2} for
a given $\omega$ in the form
\begin{equation} \label{xi_solution}
\chi_\omega (r)=r^2\left[\chi_J(\omega) J_{n+1}(\omega r)+\chi_Y(\omega) Y_{n+1}(\omega
r)\right] \,,
\end{equation}
where $\chi_J(\omega)$ and $\chi_Y(\omega)$ are some $\omega$-dependent constants that
can be chosen arbitrarily until the boundary conditions are specified, and $J_{n+1}$ and
$Y_{n+1}$ are the Bessel and Neumann functions, respectively.

The asymptotic behavior of the function $\chi_\omega (r)$ in the neighborhood of the
point $r=0$ is determined in the leading order by the asymptotic of the Neumann function:
\begin{equation} \label{asymptN}
\chi_\omega(r)\rightarrow - \frac{2^{n+1} n!\,\chi_Y(\omega)}{\pi \omega^{n+1}\,r^{n-1}}
\, , \quad r \rightarrow 0 \, .
\end{equation}
The requirement of the regularity of the solution at $r = 0$ leads to the condition
$\chi_Y(\omega) = 0$ for all modes with $n \geq 2$.

Now, the general solution of the master equation \eqref{master equation flat} with
oscillatory behavior can be written in the form of an integral over all possible values
of the parameter $\omega$:
\begin{equation} \label{master1}
\muk_J (\tau, r)= r^2 \int^{\infty}_{-\infty} d\omega\, \hat \muk_J
(\omega)\, J_{n+1}(\omega r)\,e^{i \omega \tau} \, .
\end{equation}

In the case $W\equiv - \,\omega^2 < 0$, we obtain the solution of \eqref{separation
variables 2} for a given $\omega$ in the form
\begin{equation} \label{xi_solution_tachion}
\chi_\omega(r)=r^2\left[\chi_I(\omega) I_{n+1}(\omega r)+\chi_K(\omega) K_{n+1}(\omega
r)\right] \,,
\end{equation}
where $\chi_I(\omega)$ and $\chi_K(\omega)$, again, are some $\omega$-dependent
constants, and $I_{n+1}(\omega r)$ and $K_{n+1}(\omega r)$ are the modified Bessel
functions of the first and second kind, respectively.

Similarly to the previous case, the absence of singularities at the point $r = 0$ implies
$\chi_K(\omega) = 0$ for all modes with $n \geq 2$. The general solution of the master
equation \eqref{master equation flat} with exponential behaviour can be written as:
\begin{equation} \label{master2}
\muk_I (\tau, r)= r^2 \int^{\infty}_{-\infty} d\omega\, \hat \muk_I
(\omega)\, I_{n+1}(\omega r)e^{\omega \tau} \, .
\end{equation}

The complete general solution of the Mukohyama master variable is the sum of
(\ref{master1}) and (\ref{master2}), namely
\beq \label{master3}
\muk (\tau, r)= r^2 \int^{\infty}_{-\infty} d\omega\, \hat \muk_J (\omega)\,
J_{n+1}(\omega r)\,e^{i \omega \tau} + r^2 \int^{\infty}_{-\infty} d\omega\, \hat \muk_I
(\omega)\, I_{n+1}(\omega r)e^{\omega \tau} \, .
\eeq
Here, the functions $\hat \muk_J (\omega)$ and $\hat \muk_I (\omega)$ are arbitrary and
are determined by the boundary conditions on the brane [the equations on the brane are
the boundary conditions for the master variable $\muk(\tau, r)$]. The exact analysis of
the combined brane--bulk system of integro-differential equations looks very complicated.
We can only hope to use the general solution in the bulk to find conditions under which
the effects of nonlocality can be neglected, and the dynamics of the brane perturbations
becomes closed.

\section{Perturbations of a marginally closed brane}
\label{sec: perturbations on a marginally closed brane}

Solution (\ref{master3}) for the Mukohyama master equation is obtained for the case of a
spatially closed brane, which represents a three-sphere bounding a four-ball in bulk
space. Such a configuration is very convenient because the boundary condition in the bulk
away from the brane in this case is specified uniquely just as the regularity condition
of the metric inside the ball.  At the same time, in the limit of relatively small
spatial curvature of the brane or, more precisely, under the condition
\begin{equation}\label{flatness}
a^2 H^2 \gg 1 \, ,
\end{equation}
the difference between the spatially closed and flat brane geometries is irrelevant for
the evolution on observationally significant spatial scales, i.e., for $n \gg 1$. In this
case, expressions \eqref{weyl projection rho master} and \eqref{weyl projection pi master
t} for the energy density and anisotropic stress of the Weyl fluid reduce, respectively,
to
\begin{equation}\label{rho-flat}
\frac{\delta \rho_{\C}}{m^2} = -\,\frac{n^4}{3 a^5}\,\muk_{\rm b} \,,
\end{equation}
\begin{equation} \label{pi-flat}
\frac{\delta \pi_{\C}}{m^2} = - \frac{1}{2a} \left[\ddot{\muk}_{\rm b} -
\left(H+\frac{\dot{H}}{H}\right)\,\dot{\muk}_{\rm b} + \frac{n^2}{3a^2}\, \muk_{\rm b} -
\frac{\dot{H}}{a H^2}\left(\partial_r \muk\right)_{\rm b} \right]\, .
\end{equation}

It is clear that the only term preventing the closure of these equations on the brane is
the term proportional to $\left(\partial_r \muk\right)_{\rm b}$ on the right-hand side of
\eqref{pi-flat}. Remarkably, in a marginally closed brane the contribution from this term
can be neglected resulting in the following relationship between $\delta \rho_{\C}$ and
$\delta \pi_{\C}$
\begin{equation} \label{pi-closed}
\delta \pi_{\C} = \frac{3a^4}{2n^4} \left[\delta {\ddot \rho}_{\C} + \left(9 H -
\frac{\dot{H}}{H}\right)\, \delta {\dot \rho}_{\C} + \left(20 H^2 +
\frac{n^2}{3a^2}\right)\, \delta \rho_{\C} \right]\, ,
\end{equation}
which formally closes the system of brane perturbations.

Let us explicitly demonstrate that the contribution from $\left(\partial_r
\muk\right)_{\rm b}$ is typically small and can be neglected.  First, note
that\footnote{This logic does not mean that the derivative ${(\partial_y \muk)}_\rb$ with
respect to the Gaussian normal coordinate $y$ relative to the brane can also be
neglected. In fact, for this derivative, we have ${(\partial_y \muk)}_\rb = - a H
\left(\partial_{\tau} \muk\right)_\rb - \sqrt{1+a^2 H^2}\left(\partial_{r}
\muk\right)_\rb $. In view of \eqref{rule of differentiation} and condition
\eqref{flatness}, we can relate ${(\partial_y \muk)}_\rb$ and ${(\partial_r \muk)}_\rb$
as $ {(\partial_y \muk)}_\rb \approx - \dot{\muk}_\rb - \left(\partial_{r}
\muk\right)_\rb/(aH) $. Therefore, the dimensionless quantity ${N \equiv \left(\partial_y
\muk\right)_{\rm b}/(H \muk_{\rm b}) }$ can roughly be estimated as $N \approx -
\dot{\muk}_\rb/(H \muk_{\rm b}) \sim -1$. This estimate is in agreement with the result
of \cite{Sawicki:2006jj}, where the quantity of $N$ also has relatively large negative
values during the numerical integration in frames of the dynamical-scaling approximation.
In fact, the equality $N = - \dot{\muk}_\rb/(H \muk_{\rm b})$ by itself closes the
dynamics of brane perturbations.} $\left(\partial_r \muk\right)_{\rm b} \sim \muk_\rb /
a$, while $\dot \muk_\rb \sim H \muk_\rb$ and $\ddot \muk_\rb \sim H^2 \muk_\rb$, in
which case the last term in the brackets on the right-hand side of (\ref{pi-flat}) is of
the order $\frac{\dot{H}}{a^2H^2}\muk_\rb$, which is much smaller than the sum of the
first and second terms in the brackets in a marginally closed universe satisfying $a^2H^2
\gg 1$.  We therefore conclude that the last term in the brackets of (\ref{pi-flat}) can
be neglected in view of the condition
\begin{equation} \label{condition} \left|\frac{\dot{H}}{H^2}\right| \ll a^2 H^2\, ,
\end{equation} valid in a {\em marginally\/} closed universe. Since $q = -{\ddot a}/aH^2$
defines the deceleration parameter, the above equation is equivalent to $1+q \ll a^2H^2$
(provided $q \geq -1$).

Let us provide more rigorous mathematical arguments for the smallness of
$\left(\partial_r \muk\right)_{\rm b}$. Consider first the oscillatory part
\eqref{master1} of the master variable (\ref{master3}). Using the integral representation
for Bessel functions of integer order $s$ (see, e.g., \cite{Abr_Steg}),
\begin{equation}
J_{s}(z) = \frac{(-i)^s}{\pi} \int^{\pi}_{0} d \phi \cos (s \phi) \,
e^{i z\cos\phi} \, ,
\end{equation}
we can write the master variable \eqref{master1} in the form
\begin{equation}
\muk_J (\tau, r)= \frac{(-i)^{(n+1)}}{\pi}\int^{\pi}_{0} d \phi \cos \left[ (n+1)\phi \right]
\int^{\infty}_{-\infty} d\omega\, \hat \muk_J (\omega)\, e^{\,i\, \omega \,(r \cos\phi +
\tau)}\,r^2 \,.
\end{equation}
At the brane ($r=a$), we have, for this solution,
\begin{equation}
\muk_{\rm b}(t)= \frac{(-i)^{(n+1)}}{\pi}\int^{\pi}_{0} d \phi \cos [(n+1)\phi]
\int^{\infty}_{-\infty} d\omega\, \hat \muk_J (\omega)\, e^{\,i\, \omega \,a(\cos\phi + 1)}\,a^2
\,, \end{equation}
\begin{equation} \label{master-flat}
\left(\partial_r \muk\right)_{\rm b}(t)= \frac{(-i)^{(n+1)}}{\pi}\int^{\pi}_{0} d \phi
\cos [(n+1)\phi] \int^{\infty}_{-\infty} d\omega\, \hat \muk_J (\omega)\, e^{\,i\, \omega
\,a(\cos\phi + 1)}\,\left(2 a +i\omega a^2\cos\phi\right) \, ,
\end{equation}
where in the exponent we have used the approximation $\tau (t) \approx a (t)$, valid for
a marginally closed brane ($a H \gg 1$).

The anisotropic stress \eqref{pi-flat} of the Weyl fluid can now be written in the form
\begin{equation}
\frac{\delta \pi_{\C}}{m^2} =
 -\,\frac{(-i)^{(n+1)}}{2 \pi a}\int^{\pi}_{0} d \phi \cos
[(n+1)\phi] \, \int^{\infty}_{-\infty} d\omega\, \hat \muk_J (\omega)\, e^{\,i\, \omega
\,a(\cos\phi + 1)}\, \Pi(t; \omega, \phi)\,,
\end{equation}
where
\begin{eqnarray}
\Pi(t; \omega, \phi) &=& \Pi_{00}(t)+i\,\omega \Pi_{10}(t) + i\,\omega \cos\phi\,
\Pi_{11}(t) \nonumber \\
&& {} + \omega^2 \Pi_{20}(t) + \omega^2 \cos\phi\, \Pi_{21}(t) + \omega^2 \cos^{2}\phi\,
\Pi_{22}(t) \, .
\end{eqnarray}

The functions $\Pi_{ij}(t)$ can be calculated explicitly.  We are interested here only in
$\Pi_{00}$ and $\Pi_{11}$ because the term with $\left(\partial_r \muk\right)_{\rm b}$
contributes only to these functions [see \eqref{master-flat}]. The result is
\begin{equation} \Pi_{00}(t) = (2 a
\ddot{a}+2\dot{a}^2)-2a\dot{a}\left(H+\frac{\dot{H}}{H}\right)+\frac{n^2}{3}
-\underline{2\,\frac{\dot{H}}{H^2}} =  2 a^2 H^2 + \frac{n^2}{3} -
\underline{2\,\frac{\dot{H}}{H^2}} \,,
\end{equation}
\begin{equation}
\Pi_{11}(t)= (a^2 \ddot{a}+4 a \dot{a}^2)-a^2 \dot{a}\left(H+\frac{\dot{H}}{H}\right)
-\underline{a\,\frac{\dot{H}}{H^2}} =  4 a^3 H^2 - \underline{a\,\frac{\dot{H}}{H^2}}
\,,
\end{equation}
where the underlined terms are those stemming from the term with $\left(\partial_r
\muk\right)_{\rm b}$ in \eqref{pi-flat}. They can be neglected in the overall expression
if, in addition to \eqref{flatness}, we also assume (\ref{condition}).   (Note that the
terms connected with $\ddot \muk_\rb$ and $\dot \muk_\rb$ in \eqref{pi-flat}, in general,
are not small and cannot be neglected.) Neglecting $\left(\partial_r \muk\right)_{\rm b}$
in \eqref{pi-flat} allows us to relate the anisotropic stress $\delta \pi_{\C}$ of the
Weyl fluid and its density perturbation $\delta \rho_{\C}$ through (\ref{pi-closed}).

One can use a similar integral representation for the modified Bessel functions of the
first kind to obtain the same result in the case of the exponential part (\ref{master2})
of the complete solution (\ref{master3}). This means that, under very mild conditions
\eqref{flatness} and \eqref{condition} describing a marginally closed brane, both parts
of the bulk solution \eqref{master3} lead to identical equations on the brane, namely
(\ref{pi-closed}), which effectively closes the system of brane perturbations.

Since the system of equations \eqref{con-delta1}--\eqref{w33}, supplemented by
\eqref{pi-closed}, is closed, it may be integrated (at least numerically). It is
important to note that equation \eqref{pi-closed} is valid both for sub-Hubble ($n \gg a
H$) and super-Hubble ($n \ll a H$) scales as long as $n \gg 1$. The only restrictions we
have used in the derivation of this equation are conditions \eqref{flatness} and
\eqref{condition}.

Expression \eqref{pi-closed} generalizes the result of Koyama and Maartens
\cite{Koyama:2005kd}
\begin{equation} \label{qs}
\delta \pi^{({\rm qs})}_{\C} \approx \frac{a^2}{2n^2} \delta \rho^{({\rm qs})}_{\C} \, ,
\end{equation}
which was derived for perturbations on sub-Hubble spatial scales in the quasi-static
approximation. In fact, the essence of the quasi-static approximation of
\cite{Koyama:2005kd} is the assumption that all terms with time derivatives of $\delta
\rho_\C$ can be neglected relative to those with spatial gradients.  On sub-Hubble
spatial scales ($n \gg a H$), this assumption immediately transforms \eqref{pi-closed}
into \eqref{qs}.

Proceeding further in the case of a marginally closed brane ($a H \gg 1$), and neglecting
the spatial-curvature term on the left-hand-side of the background evolution equation
(\ref{background}), one can conveniently write this equation (for the normal branch) in
the form (recall that $C=0$ and $\Lambda = 0$ in the bulk)
\begin{equation} \label{background-flat}
H^2 = \frac{\rho + \sigma}{3 m^2} + \frac{2}{\ell^2}\left[ 1 - \sqrt{1 + \ell^2
\left(\frac{\rho + \sigma}{3 m^2} \right)} \right] =  \frac{1}{\ell^2} \left[ 1 -
\sqrt{1 + \ell^2 \left(\frac{\rho + \sigma}{3 m^2} \right)} \right]^2 \, .
\end{equation}
The functions $\beta$ and $\gamma$ \eqref{beta-gamma} in this case
can be expressed in terms of $H$ and $\dot{H}$:
\begin{equation} \label{beta-flat}
\beta = \,-\, 2 \sqrt{1 + \ell^2 \left(\frac{\rho + \sigma}{3 m^2}
\right) } = - \,2\,(1+\ell H) \, ,
\end{equation}
\begin{equation} \label{gamma-flat}
3\,\gamma - 1 =\, \frac{\dot \beta}{H \beta} \,=\, \frac{\ell \dot{H}}{H (1 + \ell H)} \, .
\end{equation}

For perturbations of pressureless matter without anisotropic stress ($p = 0$, $\zeta =
0$), described in the general case by \eqref{one}--\eqref{three}, using \eqref{upsilon
C}, \eqref{pi-closed}, \eqref{beta-flat}, \eqref{gamma-flat} and proceeding to the
variable $\muk_\rb$, connected with $\delta \rho_\C$ via an algebraic relation
\eqref{rho-flat}, we derive the following {\em closed\/} system of equations on the
brane:
\begin{equation}\label{closed-1}
\ddot \Delta + 2 H \dot \Delta = \left[\frac{\ell H}{1 + \ell H}  -
\frac{\ell \dot H}{H ( 1 + \ell H)^2}
\right] \frac{\rho \Delta}{2 m^2} + \frac{n^4}{6 a^5}
\left[ \frac{\ell \dot H}{H ( 1 + \ell H)^2} + \frac{2}{1 + \ell H} \right]
\muk_\rb  \, ,
\end{equation}
\ber \label{closed-2}
\ddot \muk_\rb + \left( 2 \ell^{-1} - H - \frac{\ell \dot H}{1 + \ell H} \right) \dot
\muk_\rb + \left[ \left( 1 + \frac{\ell \dot H}{3 H (1 + \ell H)} \right) \frac{n^2}{a^2}
- 2 H \ell^{-1} - \frac{\dot H}{1 + \ell H} \right] \muk_\rb \nonumber \\ =
\frac{a^3}{n^2} \left[2 + \frac{\ell \dot{H}}{ H(1 + \ell H)}\right] \frac{\rho
\Delta}{m^2} \, . \quad
\eer

Equations \eqref{con-delta1}, \eqref{two}, \eqref{difference}, and \eqref{pi-closed} can
then be used to determine the relativistic potentials $\Phi$ and $\Psi$ in (\ref{metric})\,:
\beq \label{psi}
\Psi = - \frac{a^2}{n^2}\, \frac{\ell H}{1 + \ell H} \, \frac{\rho \Delta}{2 m^2} +
\frac{1}{2 a (1 + \ell H)} \left[ H \dot \muk_\rb - H^2 \muk_\rb - \frac{n^2}{3 a^2}
\muk_\rb \right] \, ,
\eeq
\beq \label{difference-1}
\Phi - \Psi = - \frac{1}{a \left[ \ell \dot H / H + 2 (1 + \ell H) \right]} \left[ \ddot
\muk_\rb - H \left( 1 + \frac{\dot H}{H^2} \right) \dot \muk_\rb + \frac{n^2}{3 a^2}
\muk_\rb \right] \, .
\eeq

Equations (\ref{closed-1})--(\ref{difference-1}) form one of the main results of this
paper.

\section{Perturbations of pressureless matter during the general-relativistic regime of
evolution} \label{sec: general-relativistic regime}

As we have noted at the end of Sec.~\ref{sec: perturbations on the brane}, our braneworld
model leads to three distinct effects concerning the evolution of matter perturbations:
(i)~it modifies the evolution of the Hubble parameter $H(t)$; (ii)~it produces a
time-dependent renormalization of the effective gravitational constant; and (iii)~it
introduces the gravity of the Weyl fluid. In a braneworld model with induced gravity, the
first two effects can be quite weak at early times, and it is mostly the third effect
that distinguishes this theory from general relativity. In fact we know that, in contrast
to the Randall--Sundrum model, in the induced-gravity models the effect of an extra
dimension on cosmic expansion can be negligibly small at early times. The same is true
also for the effective renormalization of the gravitational constant. But perturbations
of the Weyl fluid exist at all times and, in principle, can never be ignored.

Let us then consider the evolution of the brane under the conditions
\begin{equation}\label{condition_background-gr}
\frac{(\rho+\sigma)}{m^2} \gg \ell^{-2}\, , \qquad H \gg \ell^{-1} \, ,
\end{equation}
implying that the effect of the extra dimension (parameterized by the inverse length
$\ell^{-1} = M^3 / 2 m^2$) on the background evolution is small.  In this approximation,
equation \eqref{background-flat} turns into a `general-relativistic' expansion law
\beq \label{background-gr}
H^2\approx\frac{(\rho+\sigma)}{3m^2} \, ,
\eeq
so that
\beq \label{happrox}
\dot{H}\approx - \,\frac{\rho \,(1+w)}{2 m^2} \, , \quad \beta\approx -2\ell H \, , \quad
\gamma\approx -\frac{\left(1+3w-2\sigma/\rho\right)}{6(1+\sigma/\rho)} \, .
\eeq
Here, $w(t) \equiv p(t)/\rho(t)$ is the parameter of the equation of state of matter,
which for pressureless matter would be equal to zero. In terms of the usual cosmological
parameters \cite{Sahni:2002dx},
\beq
\Omega_\m = \frac{\rho}{3m^2H^2} \, , \quad \Omega_\sigma = \frac{\sigma}{3m^2H^2} \, ,
\quad \Omega_\ell = \frac{1}{\ell^2H^2} \, ,
\eeq
the inequalities in (\ref{condition_background-gr}) reduce to
\beq
\Omega_\m + \Omega_\sigma \gg \frac{\Omega_\ell}{3} \, , \quad \Omega_\ell \ll 1 \, .
\eeq

We are going to consider the role of perturbations
of the Weyl fluid in this regime. Although this issue can be discussed for any type of
matter, we are mostly interested in the case of pressureless matter ($w = 0$) without
anisotropic stress ($\zeta = 0$). In this case, the system of equations \eqref{closed-1},
\eqref{closed-2} simplifies to
\begin{equation}\label{pressureless gr delta sigma}
\ddot \Delta + 2 H \dot \Delta = \frac{\rho \Delta}{2 m^2} + \frac{n^4}{ 6
a^5 \ell H} \left( 2 + \frac{\dot H}{H^2} \right) \muk_\rb \, ,
\end{equation}
\begin{equation} \label{pressureless gr weyl sigma}
\ddot \muk_\rb - H \left( 1 + \frac{\dot H}{H^2} \right) \dot \muk_\rb + \frac{n^2}{a^2}
\left( 1 + \frac{\dot H}{3 H^2} \right) \muk_\rb = \frac{a^3}{n^2} \left( 2 + \frac{\dot
H}{H^2} \right) \frac{\rho \Delta}{m^2}  \, ,
\end{equation}
where in \eqref{pressureless gr delta sigma} we neglected terms proportional to $1 / \ell
H$ with respect to unity, according to \eqref{condition_background-gr}. Up to this
accuracy, the effect of renormalization of the gravitational coupling can be ignored.  In
this case, using (\ref{pressureless gr weyl sigma}), equations (\ref{psi}) and
\eqref{difference-1} are simplified to
\beq \label{psi2}
\Psi = - \frac{a^2}{n^2}\, \frac{\rho \Delta}{2 m^2} + \frac{1}{2 \ell a} \left[ \dot
\muk_\rb - H \muk_\rb - \frac{n^2}{3 a^2 H} \muk_\rb \right] \, ,
\eeq
\beq \label{difference-2}
\Phi - \Psi = \frac{1}{\ell H} \left( \frac{n^2}{3 a^3} \muk_\rb - \frac{a^2}{n^2}
\frac{\rho \Delta}{m^2} \right) = - \frac{a^2}{n^2 \ell H}\, \frac{ \rho \Delta + \delta
\rho_\C}{m^2} \, .
\eeq

Choosing the scale factor $a$ as a new variable instead of the time $t$, one writes
equations \eqref{pressureless gr delta sigma}, \eqref{pressureless gr weyl sigma} as
follows:
\begin{equation}\label{hoho1}
a^2 \Delta'' + \frac{3a}{2} \left(\frac{\rho + 2 \sigma}{\rho + \sigma}\right) \Delta' -
\frac{3 \rho}{2 (\rho + \sigma)} \Delta =  \frac{n^4}{12\, \ell \, a^5 H^3}
\left(\frac{\rho + 4 \sigma}{\rho + \sigma}\right) \muk_{\rm b} \, ,
\end{equation}
\begin{equation}\label{hoho2}
a^2 {\muk}_{\rm b}'' + \frac{n^2}{2 a^2H^2} \left(\frac{\rho + 2 \sigma}{\rho +
\sigma}\right) \muk_{\rm b} = \, \frac{3 a^3}{2n^2} \left(\frac{\rho + 4 \sigma}{\rho +
\sigma}\right) \left( \frac{\rho}{\rho + \sigma}\right) \Delta \, ,
\end{equation}
where a prime denotes the derivative with respect to $a$.

In what follows, we always assume conditions (\ref{condition_background-gr}) to be valid,
so that braneworld effects do not contribute substantially to the expansion of the
universe which is effectively described by a marginally closed $\Lambda$CDM, evolving
according to (\ref{background-gr}).
(Note that the brane tension $\sigma$ plays the role of the cosmological constant on the
brane.) As we shall show, bulk effects, in the form of perturbations of dark radiation,
alter the growth of density perturbations in brane-based $\Lambda$CDM relative to those
in standard $\Lambda$CDM. We shall examine two complementary regimes of the effective
$\Lambda$CDM cosmology: (i)~the matter dominated epoch when $\rho \gg \sigma$, and
(ii)~de~Sitter-like expansion, when $\rho \ll \sigma$.

\subsection{Matter dominated epoch}
\label{sec:matter_domination}

An important exact solution of this system of equations can be found when, during early
times, the matter density dominates over the brane tension. Under the condition $\rho \gg
\sigma$, equations \eqref{hoho1}, \eqref{hoho2} simplify, respectively, to
\begin{equation}\label{gr_no_sigma_delta}
a^2 \Delta'' + \frac{3a}{2} \Delta' - \frac{3}{2}\Delta = \frac{n^4}{12\, \ell \, a^5
H^3} \muk_{\rm b} \, ,
\end{equation}
\begin{equation}\label{gr_no_sigma_weyl}
a^2 {\muk}_{\rm b}'' + \frac{n^2}{2 a^2H^2}\, \muk_{\rm b} = \, \frac{3 a^3}{2n^2} \Delta
\, .
\end{equation}

Na\"{\i}vely, the first term in \eqref{gr_no_sigma_weyl} can be estimated as $a^2
{\muk}_{\rm b}'' \sim {\muk}_{\rm b}$, implying that it can be neglected if $n\gg aH$.
This rough estimate leads to the quasi-static approximation of Koyama and Maartens on
sub-Hubble scales. Equations \eqref{gr_no_sigma_delta}, \eqref{gr_no_sigma_weyl}
 in this approximation take the form
\begin{equation}\label{gr_no_sigma_weyl KM}
{\muk}_{\rm b} \approx \, \left(\frac{3 a^5 H^2}{n^4}\right) \Delta \,,
\end{equation}
\begin{equation}\label{gr_no_sigma_delta KM}
a^2 \Delta'' + \frac{3 a}{2} \Delta' - \frac{3}{2}\Delta \approx \frac{\Delta}{4\ell H}
\, .
\end{equation}
The impact of the Weyl fluid on the evolution of matter perturbations  is described by
the term on the right-hand side of \eqref{gr_no_sigma_delta KM}. Note that the system of
equations \eqref{pressureless gr delta sigma}, \eqref{pressureless gr weyl sigma} was
derived under the assumption that $\ell H \gg 1$ in which case $\Delta/\ell H \ll
\Delta$. In other words the impact of the Weyl fluid is of the same order as the terms
neglected while deriving \eqref{pressureless gr delta sigma} and can safely be neglected
in \eqref{gr_no_sigma_delta KM}. We, therefore, conclude that, {\em in the
general-relativistic regime}, the Koyama--Maartens solution predicts only an
insignificantly small correction to the general-relativistic equation of cosmological
perturbations
\begin{equation}
a^2 \Delta'' + \frac{3 a}{2} \Delta' - \frac{3}{2}\Delta = 0~.
\end{equation}
This is in agreement with the numerical analysis of \cite{Cardoso:2007xc}.

One should not forget, however, that neglecting the term with the second derivative in
\eqref{gr_no_sigma_weyl} {\em reduces the order\/} of that differential equation. In
fact, the system of equations \eqref{gr_no_sigma_delta} and \eqref{gr_no_sigma_weyl} can
easily be transformed into a single differential equation of the fourth order for the
function $\Delta$. Thus, in general, we expect to obtain four general solutions for
$\Delta$, while the quasi-static approximation gives only two of them.  To perform a more
careful analysis, it is convenient to introduce a new variable
\begin{equation}\label{y definition}
x \equiv \sqrt{2}\left(\frac{n}{aH}\right) \, ,
\end{equation}
such that the domain $x\gg 1$ corresponds to the mode in the sub-Hubble regime, while
$x\ll 1$ is a super-Hubble limit. Using the law of cosmological evolution
$H^2\approx{a_m}/{a^3}$  with constant $a_m$, which is valid at early times (during
matter domination) when braneworld effects can be neglected, one can write the system of
equations \eqref{gr_no_sigma_delta}, \eqref{gr_no_sigma_weyl} in terms of the new
variable $x$:
\begin{equation}\label{gr_no_sigma_delta_dimensionless_2}
x^2 \Delta'' + 2x \Delta' - 6 \Delta =\frac{1}{x^2}\frac{d}{d x}\left[x^6\frac{d}{d
x}\left(\frac{\Delta}{x^2}\right)\right] = P(x) \, ,
\end{equation}
\begin{equation}\label{gr_no_sigma_weyl_dimensionless_2}
P'' + \frac{1}{x}\,P' + \left(1-\frac{1}{x^2}\right)P = \,\frac{\Delta}{\ell H} \,,
\end{equation}
where the prime denotes differentiation with respect to $x$,
\begin{equation}
P(x) \equiv \frac{\sqrt{2}\,n^5\,\muk_{\rm b} (x)}{3a_m^2\ell x} \, ,
\end{equation}
and we have used the relation
\begin{equation}
\frac{a_m \,x^3}{2 \sqrt{2}\, \ell \,n^3} = \frac{1}{\ell H} \, .
\end{equation}

One can easily verify that the system \eqref{gr_no_sigma_delta_dimensionless_2} \&
\eqref{gr_no_sigma_weyl_dimensionless_2} is equivalent to a single fourth-order
differential equation for $\Delta(x)$:
\begin{equation}\label{gr_no_sigma_delta_dimensionless_fourth_order}
x^2 \Delta^{(4)} +7 x \Delta^{(3)} + (3 + x^2) \Delta'' + 2x \left(1-\frac{3}{x^2}\right)
\Delta' - 6 \left(1-\frac{1}{x^2}\right) \Delta = \frac{\Delta}{\ell H} \,.
\end{equation}
It is clear that, in view of \eqref{condition_background-gr}, the expression on the
right-hand side of the last equation is negligibly small compared to the term
proportional to $\Delta$ on the left-hand side. The same is also true for the expression
on the right-hand side of \eqref{gr_no_sigma_weyl_dimensionless_2}, which should be
omitted in our approximation. Since $P \propto \muk_{\rm b}$, one can conclude that, in
the general-relativistic regime, the back reaction from $\Delta$ on the evolution of the
perturbations of the Weyl fluid, $\muk_{\rm b}$,
 can be neglected, which
facilitates finding solutions during the matter dominated regime.

We therefore arrive at the following important result: neglecting $\Delta/\ell H$ in
\eqref{gr_no_sigma_weyl_dimensionless_2} reduces that equation to a homogeneous equation
in the dark-radiation variable $P \propto \muk_{\rm b}$. This equation can be solved
exactly and then substituted back into \eqref{gr_no_sigma_delta_dimensionless_2} to
provide an equation for the density contrast on the brane, $\Delta$, which is the
quantity that interests us most. Proceeding in this manner and omitting the inhomogeneous
part of \eqref{gr_no_sigma_weyl_dimensionless_2}, one finds its general solution to be
\begin{equation}\label{P homogeneous solution}
P(x) = F J_1(x) + G Y_1(x)\, ,
\end{equation}
where $F$ and $G$ are constants, and $J_1(x)$ and $Y_1(x)$ are Bessel and Neumann
functions of the first order, respectively. Using \eqref{P homogeneous solution}, one can
integrate \eqref{gr_no_sigma_delta_dimensionless_2} and express its general solution in
the form:
\begin{equation}\label{Delta general solution_h}
\Delta(x)= A x^2 + \frac{B}{x^3} - F h_J(x)-G h_Y(x) \, ,
\end{equation}
where $A$ and $B$ are constants, corresponding to the homogeneous solution of
\eqref{gr_no_sigma_delta_dimensionless_2}, and the functions $h_J(x)$ and $h_Y(x)$ are
\begin{equation}\label{h_j}
h_J(x) = \frac{J_2(x)}{15x}\left(x^4\left[\frac{\pi}{2}H_1(x)-1\right]-x^2+3\right)+
\frac{J_1(x)}{45}\left(x^4\left[1-\frac{3\pi}{2x}H_2(x)\right]+3x^2+9\right)\,,
\end{equation}
\begin{equation}\label{h_x}
h_Y(x) = \frac{Y_2(x)}{15x}\left(x^4\left[\frac{\pi}{2}H_1(x)-1\right]-x^2+3\right)+
\frac{Y_1(x)}{45}\left(x^4\left[1-\frac{3\pi}{2x}H_2(x)\right]+3x^2+9\right)\,,
\end{equation}
with $H_{1}(x)$ and $H_{2}(x)$ being Struve functions of first and second order,
respectively.

Note that $x \propto (aH)^{-1}$ and, in a matter dominated universe, $aH \propto
a^{-1/2}$, so that the first ($\propto x^2 \propto a \propto t^{2/3}$) and second
($\propto x^{-3} \propto a^{-3/2} \propto t^{-1}$) terms in \eqref{Delta general
solution_h} describe the evolution of growing and decaying modes, respectively,
of perturbations in general relativity without the Weyl fluid. The contribution from the
Weyl fluid is given by the last two terms in \eqref{Delta general solution_h}, involving two additional constants of
integration $F$ and $G$.

We now use these results to obtain the behaviour of the relativistic potentials $\Psi$
and $\Phi$ in (\ref{metric}). To the zero-order approximation in the small quantity $1 /
\ell H$, used in this section, we find from \eqref{psi2} and \eqref{difference-2}
\beq \label{psi-1}
\Psi = - \frac{3}{x^2} \Delta (x) + \frac{3}{x^2} \left( \frac{P (x)}{x} \right)' -
\frac{P (x)}{x^2} \, ,
\eeq
\beq \label{diff-1}
\Phi - \Psi = \frac{2}{x^2} P (x) \, .
\eeq

The asymptotic evolution of the modes $P$ and $\Delta$ prior to Hubble-radius crossing is
given by the leading asymptotic expansion of \eqref{P homogeneous solution} and
\eqref{Delta general solution_h} in the domain $x\ll 1$\,:
\ber
P (x) &\approx& F \left( \frac{x}{2} - \frac{x^3}{16} \right)
- G \left[ \frac{2}{\pi x} - \frac{x \log x}{\pi} \right] \, , \\
\Delta(x) &\approx& A x^2 + \left(B+\frac{4G}{5\pi}\right)\frac{1}{x^3} - \frac{F}{8}\, x
+ \frac{G}{3\pi x} \, , \label{Delta general solution super-Hubble}
\eer
where we remind the reader that $x = \sqrt{2}n/aH$; see (\ref{y definition}). The first
and third terms in $\Delta (x)$ correspond to growing modes, and the second and fourth to
decaying modes. Note that the main correction to the general-relativistic result is given
by the third term in \eqref{Delta general solution super-Hubble}, which {\em grows with
time\/} as $x \propto \sqrt{a}$.

If the mode crosses the Hubble radius, then its asymptotic behavior after that is
described by \eqref{P homogeneous solution} and \eqref{Delta general solution_h} in the
domain $x \gg 1$\,:
\ber
P (x) &\approx& \sqrt{\frac{2}{\pi x}} \left[ F \sin \left( x - \frac{\pi}{4} \right) - G
\cos \left( x - \frac{\pi}{4} \right) \right] \, , \\
\Delta(x) &\approx& \left(A-\frac{F}{15}\right) x^2 + \frac{B}{x^3} - \sqrt{\frac{2}{\pi
x^5}}\left[F \sin \left(x - \frac{\pi}{4} \right) - G \cos \left( x - \frac{\pi}{4}
\right) \right] \, . \label{Delta general solution sub-Hubble}
\eer
As we can see, the general-relativistic result on sub-Hubble scales ($\Delta \propto x^2
\propto a$) is corrected by a rapidly oscillating term whose amplitude decreases as
$x^{-5/2} \propto a^{-5/4}$. Note that, in the general-relativistic regime of the
braneworld model, corresponding to a matter-dominated universe, the oscillatory term in
\eqref{Delta general solution sub-Hubble} presents the main correction to conventional
general-relativistic behavior. This term may be insignificant for a wide class of initial
conditions [as is true for the second, decaying term ($\propto x^{-3} \propto a^{-3/2}$)
in general-relativistic cosmology]. In such a case, the solution \eqref{Delta general
solution sub-Hubble} would correspond to the quasi-static approximation of Koyama and
Maartens on sub-Hubble spatial scales.

As one can see from \eqref{Delta general solution super-Hubble} and \eqref{Delta general
solution sub-Hubble}, an important effect of the Weyl fluid is to `renormalize' the
amplitude of the usual growing ($\propto x^2 \propto a$) and decaying ($\propto x^{-3}
\propto a^{-3/2}$) modes after Hubble-radius crossing. This renormalization depends on
the primordial power spectrum, more specifically, on the amplitudes $F$ and $G$ of the
new growing and decaying modes in \eqref{Delta general solution super-Hubble}. Thus the
amplitude of the standard growing mode ($\propto x^2$) changes from $A$ at early times
($x \ll 1$), to $A-F/15$ at late times ($x \gg 1$), with $A$ being the usual
general-relativistic contribution while $F$ is the contribution from the Weyl fluid. If
decaying modes can be neglected, then the interplay is between the two amplitudes $A$ and
$F$. Since the general-relativistic growing mode ($\propto x^2$) grows at a rate
different from that of the braneworld contribution ($\propto x$), one can always
introduce the (scale dependent) parameter $x_0$ such that $|A| x^{2}_{0}= |F| x_0/8$,
which would correspond to the time when the amplitudes of the two growing modes in
\eqref{Delta general solution super-Hubble} were equal in magnitude.\footnote{Note that,
as we approach the past singularity, $x \to 0$, it is generally the new $F$-mode that
dominates in \eqref{Delta general solution super-Hubble}. Also note that the parameter
$x_0$ may well fall beyond the matter-dominated domain in a realistic cosmology involving
both matter and radiation. In this case, it still has the useful meaning as a parameter
of extrapolation, characterizing the amplitude of perturbations.} Then, after
Hubble-radius crossing, the oscillatory mode in \eqref{Delta general solution sub-Hubble}
decays, while the amplitude of the main growing mode is renormalized by the factor $(1 +
8 \zeta x_0/15)$ with $|\zeta| = 1$. If $x_0 \ll 1$ [where, in fact, we can use
asymptotics \eqref{Delta general solution super-Hubble}], then the effect of
renormalization is weak. Nevertheless, since $x_0$ can be scale dependent, the effect of
renormalization demonstrates how the presence of the Weyl fluid in our braneworld model
can affect the evolution of the power spectrum.

\subsection{The de~Sitter regime of accelerated expansion}
\label{sec:deSitter}

Let us now proceed to the regime of accelerated expansion, when $\rho \ll \sigma$, still
assuming that $\ell$ is so large that equations
\eqref{condition_background-gr}--\eqref{happrox} are satisfied.  In other words, the
late-time acceleration of our braneworld on its normal branch is supported mainly by the
brane tension, and the universe is approximately de~Sitter space.

Equations \eqref{hoho1} and \eqref{hoho2} in this asymptotic case are written,
respectively, as
\begin{equation}\label{hohoas1}
a^2 \Delta'' + 3 a \Delta' - \frac{3 \rho}{2 \sigma} \Delta =  \frac{n^4}{3\, \ell \, a^5 H^3}
\muk_{\rm b} \, ,
\end{equation}
\begin{equation}\label{hohoas2}
a^2 {\muk}_{\rm b}'' + \frac{n^2}{a^2H^2} \muk_{\rm b} = \, \frac{6 a^3 \rho}{n^2 \sigma}
\Delta \, .
\end{equation}

We can solve this system iteratively in the small quantity $\rho / \sigma \ll 1$.  In the
zero-order approximation, one can neglect the right-hand side in \eqref{hohoas2}.
Introducing the dimensionless function
\beq
Q  = \frac{H \muk_\rb}{\ell a } \, ,
\eeq
we have the following solution\,:
\beq \label{Q-1}
Q (x) = C_1 \cos x + C_2 \sin x \, ,
\eeq
where now we denote $x = n / a H$.

Then, neglecting the last term on the left-hand side of \eqref{hohoas2}, we obtain a
solution for $\Delta$:
\ber \label{Delta-1}
\Delta &=& \frac{C_1}{3 }  \left[ \cos x + 3 x \sin x + (2 - x^2) \cos x \right]
\nonumber \\ && {} + \frac{C_2}{3}  \left[ \sin x - 3 x \cos x + (2 - x^2) \sin x \right]
\nonumber \\ && {} + C_3 + C_4 x^2 \, .
\eer
We see that, on sub-Hubble spatial scales, corresponding to $x \gg 1$, the matter
perturbation $\Delta$ has an oscillatory part with decaying amplitude, whereas on
super-Hubble spatial scales ($x \ll 1$), $\Delta$ tends to a constant, in agreement with
the numerical simulations in \cite{Cardoso:2007xc}.

For the relativistic potentials, we have, similarly to (\ref{psi-1}) and (\ref{diff-1}),
\beq
\Psi = - \frac{a^2}{n^2}\, \frac{\rho \Delta}{2 m^2} + \frac{1}{2 \ell a} \left[ \dot
\muk_\rb - H \muk_\rb - \frac{n^2}{3 a^2 H} \muk_\rb \right] = - \frac{\rho (x) \Delta
(x)}{2 m^2 x^2 H^2} - \frac12 \left[ x Q' (x) + \frac{x^2}{3} Q (x) \right] \, ,
\eeq
\beq
\Phi - \Psi = \frac{1}{\ell H} \frac{n^2}{3 a^3} \muk_\rb = \frac{x^2}{3} Q (x) \, ,
\eeq
where $\rho (x) = C_0 x^3 / n^3$ with some $n$-independent constant $C_0$, and $Q (x)$
and $\Delta (x)$ are given by (\ref{Q-1}) and (\ref{Delta-1}), respectively.

The constants $C_1$--$C_4$ are related to the constants $A$, $B$, $F$, $G$ describing the
matter-dominated regime, and are determined by the initial conditions.  The initial power
spectra for primordial perturbations of matter and the Weyl fluid should be determined by
the process of their generation. We plan to examine this issue in a future work.

\section{Conclusions}
\label{sec: conclusion}

This paper proposes a novel approach to the problem of cosmological perturbations in a
braneworld model with induced gravity.  We consider a spatially closed brane that bounds
the interior four-ball of the bulk space. The background cosmological evolution on the
brane in this case is described by the normal branch for which the late-time expansion of
the universe can become phantom-like ($w_{\rm eff} \leq -1$). For our braneworld model,
the boundary conditions in the bulk become the regularity conditions for the metric
everywhere inside the four-ball. In this approach, there is no spatial infinity in the
bulk space as the spatial section is compact, which considerably simplifies the issue of
boundary conditions, reducing them simply to regularity conditions in the bulk.

We have thoroughly investigated the case where the bulk cosmological constant vanishes so
that the unperturbed bulk configuration is simply a flat space.  In this case, using the
Mukohyama master variable \cite{Mukohyama:2000ui, Mukohyama:2001yp}, we have derived a
closed system of integro-differential equations for scalar perturbations of matter and
the Weyl fluid (dark radiation). We argue that the effects of nonlocality of the dynamics
on the brane may be ignored if the brane is only marginally closed, so that the
conditions \eqref{flatness} and \eqref{condition} are satisfied. In this case, there
arises an approximate relation \eqref{pi-closed} that closes the system of equations for
perturbations on the brane.  Perturbations in a universe dominated by {\em
pressureless\/} matter are then described by the system of second-order differential
equations \eqref{closed-1} and \eqref{closed-2}, which are suitable for further
analytical or numerical integration.

We also present an exact solution (valid on all scales) of this system of equations for
the model in the matter-dominated and de~Sitter general-relativistic regime when $H \gg
\ell^{-1}$ [Eq.~\eqref{Delta general solution_h}].  Apart from the usual modes with the
quasi-static behavior previously discussed by Koyama and Maartens \cite{Koyama:2005kd},
we find two additional modes that behave monotonically before Hubble-radius crossing [see
Eq.~\eqref{Delta general solution super-Hubble}] and exhibit rapid oscillations with
decaying amplitude after Hubble-radius crossing [see Eq.~\eqref{Delta general solution
sub-Hubble}].  One of the interesting effects of the presence of the Weyl fluid is the
renormalization of the amplitude of the usual growing mode.  This might lead to a
modification of the primordial power spectrum of matter perturbations.

Our method goes beyond the quasi-static approximation and makes it possible to obtain a
complete and closed system of equations describing the growth of cosmological
perturbations for the normal branch of the braneworld model with induced gravity and
exhibiting late-time acceleration. Its main advantage is that it does not rely on any
simplifying assumptions, apart from requiring that the brane be closed and that its
spatial curvature be small, in agreement with recent CMB results \cite{Planck_2013}.

The results of sections \ref{sec:matter_domination} and \ref{sec:deSitter} indicate that
perturbative effects can provide a {\em smoking gun\/} signature of braneworld cosmology
by distinguishing the latter from $\l$CDM, even when the expansion histories of the two
models are {\em identical\/}. This takes place when $\Omega_\m+\Omega_\sigma \gg
\Omega_\ell$. In the complementary case, when $\Omega_\m\sim \Omega_\sigma \sim
\Omega_\ell$, the effective equation of state of dark energy becomes phantom-like $w_{\rm
eff} < -1$. Gravitational instability in this region of parameter space of the model will
form the subject of future investigations.

\section*{Acknowledgments}

The authors acknowledge support from the India-Ukraine Bilateral Scientific Cooperation
programme.  The work  of A.V. and Yu.S. was also partially supported by the SFFR of
Ukraine Grant No.~F53.2/028.


\end{document}